\documentclass[usegraphicx,usenatbib,letterpaper]{emulateapj}

\usepackage{ulem}
\usepackage{color}
\usepackage{graphics,graphicx}
\usepackage{times} 
\usepackage{amssymb}
\usepackage{amsmath}
\usepackage{natbib}
\usepackage{url}
\usepackage{multirow}
\usepackage{ctable}
\usepackage{threeparttable}
\newif\ifAMStwofonts
\AMStwofontstrue

\def\xmm{{\it XMM-Newton}}

\def\chandra{{\it Chandra}}
\def\swift{{\it Swift}}

\def\epicmos1{{EPIC-MOS1}}
\def\epicmos2{{EPIC-MOS2}}
\def\epicmos{{EPIC-MOS}}

\def\nustar{{\it NuSTAR}}

\def\H0{{\rm ~km~s^{-1}~Mpc^{-1}}}

\def\kev{\hbox{\rm keV}}

\def\ergps{\hbox{erg~s$^{-1}$}}

\def\msun{\hbox{$\rm M_{\odot}$}}

\def\gpcm3{g cm$^{-3}$}

\def\flx2xsp{\rm{\small FLX2XSP}}

\def\grid25{\hbox{\rm{\small GRID25}}}

\def\eg{{\it e.g.}}
\def\etc{{\it etc.}}

\def\la{\mathrel{\hbox{\rlap{\hbox{\lower4pt\hbox{$\sim$}}}{\raise2pt\hbox{$<$}}}}}
\def\ga{\mathrel{\hbox{\rlap{\hbox{\lower4pt\hbox{$\sim$}}}{\raise2pt\hbox{$>$}}}}}

\def\d25{D$_{25}$}

\def\.25{0.25 keV\thinspace}

\def\mbh{\rm $M_{\rm BH}$}

\def\ngc{NGC\,5907 ULX1}
\def\period{80}
\def\nsimsa{10,000}
\def\nchance{none}
\def\nsimsb{1000}
\def\errsig{1}
\def\p13{P13 in NGC\,7793}

\shorttitle{A 78 Day Period in NGC\,5907 ULX1}
\shortauthors{D.~J. Walton et al.}

\begin{document}

\title{A 78 Day X-ray Period Detected from NGC\,5907 ULX1 by \textit{Swift}}

\author{D. J. Walton\altaffilmark{1,2},
F. F{\"u}rst\altaffilmark{2},
M. Bachetti\altaffilmark{3},
D. Barret\altaffilmark{4,5},
M. Brightman\altaffilmark{2},
A. C. Fabian\altaffilmark{6},
N. Gehrels\altaffilmark{7},
F. A. Harrison\altaffilmark{2}, \\
M. Heida\altaffilmark{2},
M. J. Middleton\altaffilmark{6},
V. Rana\altaffilmark{2},
T. P. Roberts\altaffilmark{8},
D. Stern\altaffilmark{1},
L. Tao\altaffilmark{2,9},
N. Webb\altaffilmark{4,5}
}
\affil{
$^{1}$ Jet Propulsion Laboratory, California Institute of Technology, Pasadena, CA 91109, USA \\
$^{2}$ Space Radiation Laboratory, California Institute of Technology, Pasadena, CA 91125, USA \\
$^{3}$ INAF/Osservatorio Astronomico di Cagliari, via della Scienza 5, I-09047 Selargius (CA), Italy \\
$^{4}$ Universite de Toulouse, UPS-OMP, IRAP, Toulouse, France \\
$^{5}$ CNRS, IRAP, 9 Av. colonel Roche, BP 44346, F-31028 Toulouse cedex 4, France \\
$^{6}$ Institute of Astronomy, University of Cambridge, Madingley Road, Cambridge CB3 0HA, UK \\
$^{7}$ NASA Goddard Space Flight Center Greenbelt, MD 20771, USA \\
$^{8}$ Centre for Extragalactic Astronomy, Department of Physics, Durham University, South Road, Durham DH1 3LE, United Kingdom \\
$^{9}$ Center for Astrophysics, Tsinghua University, Beijing 100084, China \\
}

\begin{abstract}
We report the detection of a $78.1\pm0.5$ day period in the X-ray lightcurve of
the extreme ultraluminous X-ray source \ngc\
($L_{\rm{X,peak}}\sim5\times10^{40}$\,\ergps), discovered during an extensive
monitoring program with \swift. These periodic variations are strong, with the
observed flux changing by a factor of $\sim$3--4 between the peaks and the
troughs of the cycle; our simulations suggest that the observed periodicity is
detected comfortably in excess of 3$\sigma$ significance. We discuss possible
origins for this X-ray period, but conclude that at the current time we cannot
robustly distinguish between orbital and super-orbital variations.
\end{abstract}


\section{Introduction}

\ngc\ is a remarkable member of the ultraluminous X-ray source (ULX) population.
At a distance of $\sim$13.4\,Mpc, it exhibits an extreme peak X-ray luminosity of
$\sim$5$\times10^{40}$\,\ergps\ (\citealt{WaltonULXcat, Sutton13}). Its hard
X-ray spectrum below 10\,\kev\ had previously led to speculation that it might
host an intermediate mass black hole accreting in the low/hard state, similar to
what is seen in Galactic black hole binaries at low luminosities
(\citealt{Sutton12}; see \citealt{Remillard06rev} for a review of accretion states
in Galactic binaries). Our recent coordinated observations with the \nustar\ and
\xmm\ observatories have subsequently revealed a broadband X-ray spectrum
inconsistent with this identification (\citealt{Walton15}), similar to the other ULX
systems observed by \nustar\ to date (\citealt{Bachetti13, Walton14hoIX,
Walton15hoII, Rana15, Mukherjee15}). Instead, the broadband spectrum implies
instead that \ngc\ is likely a system accreting at high- or even super-Eddington
rates, as suggested by \cite{Sutton13}.

The most remarkable aspect of these \nustar\ and \xmm\ observations, however,
is the fact that we witnessed a rise in flux of $\sim$2 orders of magnitude or more
in the mere 4 days between our two observing epochs. \ngc\ was essentially
undetected in our first observation, with an implied luminosity of
$L_{\rm{X}}\lesssim2\times10^{38}$\,\ergps, before the source returned to a more
typical brightness of $L_{\rm{X}}\sim10^{40}$\,\ergps\ in our second
(\citealt{Walton15}). This event prompted us to begin a monitoring campaign with
the \swift\ observatory (\citealt{SWIFT}) in order to investigate whether this
behaviour was a common occurence. Although these observations have not
revealed such extreme variations again, here we report on the detection of an
$\sim$\period\ day periodicity in the \swift\ lightcurve.

\begin{figure*}
\hspace*{-0.6cm}
\epsscale{1.13}
\plotone{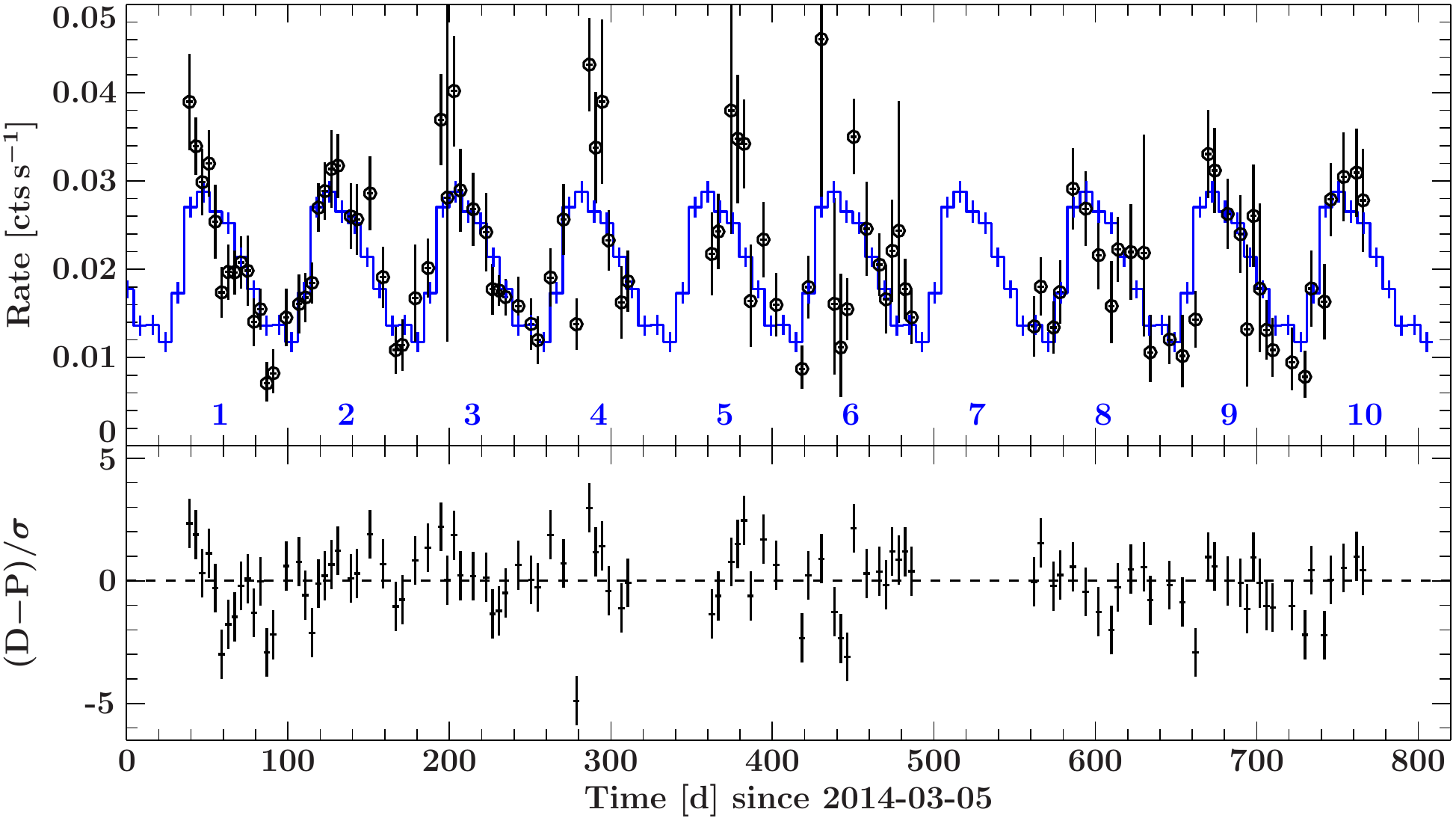}
\caption{
The \swift\ XRT lightcurve of \ngc\ obtained with our monitoring campaign, shown
with 4d bins (\textit{top panel}). A strong $78.1\pm0.5$d period is visibly present. In
addition to the XRT data, we overlay the average cycle profile in blue, and label the
individual cycles covered by the duration of our program. We also show the
agreement between the data and the average cycle profile, evaluated as the
$n\sigma$ deviation between the data ($D$) and the cycle prediction ($P$;
\textit{bottom panel}). The agreement is generally very good.
}
\vspace{0.2cm}
\label{fig_lc}
\end{figure*}

\pagebreak
\section{Swift Monitoring}
\label{sec_mon}

We began monitoring \ngc\ with \swift\ on 2014-04-14, observing every 2 days for
an initial period of 2 months. This cadence was motivated by the timescale of the
variability observed between the first two \nustar\ epochs. We then transitioned to
a longer-term monitoring program observing the source roughly every week. Aside
from a few moderately brief gaps in the coverage, this program has continued up
until the time of writing (April 2016, a duration of $\sim$700d), with an average
exposure per observation of $\sim$2\,ks. Figure \ref{fig_lc} shows the long-term
lightcurve obtained with the XRT (\citealt{SWIFT_XRT}) from these observations
with the standard \swift\ processing pipeline (\citealt{Evans09}). A strong periodicity
on the order of $\sim$\period\ days is visibly present, with the observed count rate
varying by a factor of $\sim$3--4 from peak to trough. The troughs correspond to
an observed X-ray luminosity of $\sim$10$^{40}$\,\ergps.

To search for periodicities in the data we used the epoch-folding approach, as it is
independent of the sampling of the data \citep{Leahy87}. We tested 920 trial
periods between 30\,d and 140\,d on a linearly sampled grid. The period range is set
by our requirement that each cycle be covered by at least 4 observations during the
lower cadence portion of our program, and that the duration of our monitoring
program would cover at least 5 cycles (see \citealt{Vaughan16}). To evaluate the
likelihood that at any given period the folded profile deviates significantly from the
null-hypothesis of a flat profile, we used the L-statistic as described by
\cite{Davies90}. The L-statistic is advantagous over standard $\chi^2$-statistics for
small sample sizes, as it takes the number of data points in the lightcurve and in each
phase bin of the folded profile into account. At each trial period the lightcurve was
folded into 10 phase-bins, ensuring that each bin was averaged over at least 5 data
points. The results of this search are shown in Figure \ref{fig_search}; a strong peak
in L-stat is seen at a period of 78.1 days. For reference, the minimum of the
cycle is observed at MJD 56663.0.

In addition to the XRT lightcurve, Figure \ref{fig_lc} also shows an overlay of the
average profile for the cycle, which broadly appears to resemble a fast rise,
exponential decay (FRED) profile (the rise time in the average profile is $\sim$30\%
of the cycle duration, and the decay time $\sim$50\%). This matches the data well
throughout the entire monitoring campaign to date (Figure \ref{fig_lc}, bottom panel),
but particularly so towards the beginning and the end. There appears to be a visual
indication that the period may have drifted slightly during the central portion of our
program (for example, the peak of the 5th cycle observed at $\sim$380d appears to
be slightly late). Indeed, if we re-run the period search excluding this central portion,
the significance of the period increases slightly (peak L-stat increases from
$L$=11.3 to 15.0).

\section{Significance Simulations}
\label{sec_sims}

To assess the significance of the apparent periodicity, we simulate a set of \nsimsa\
lightcurves with a red-noise power spectrum, typical of accreting black holes
(\eg\ \citealt{Vaughan03}), following the method of \cite{Timmer95}. More precisely,
we assume the power spectrum follows a simple powerlaw with a slope of
$\alpha=2.0$. These lightcurves were initially simulated with a time resolution of
2\,ks, and a total continuous duration $\sim$20 times longer than the real \swift\
lightcurve, to account for red-noise leakage on the timescales of interest. From
each of these initial lightcurves we selected a random segment matching the
duration of the \swift\ monitoring, and drew data from this segment with a cadence
that broadly matched the real observations (`observation' times were randomised 
by $\pm$1 day in comparison to the real lightcurve, to mimic the randomness of a
realistic observing process). The resulting lightcurves were then normalised to
match the average countrate and variance observed in the real data; counting
statistics were adopted for the statistical uncertainties on each bin in the final
simulated lightcurves. 

\begin{figure*}
\hspace*{-0.6cm}
\epsscale{1.13}
\plotone{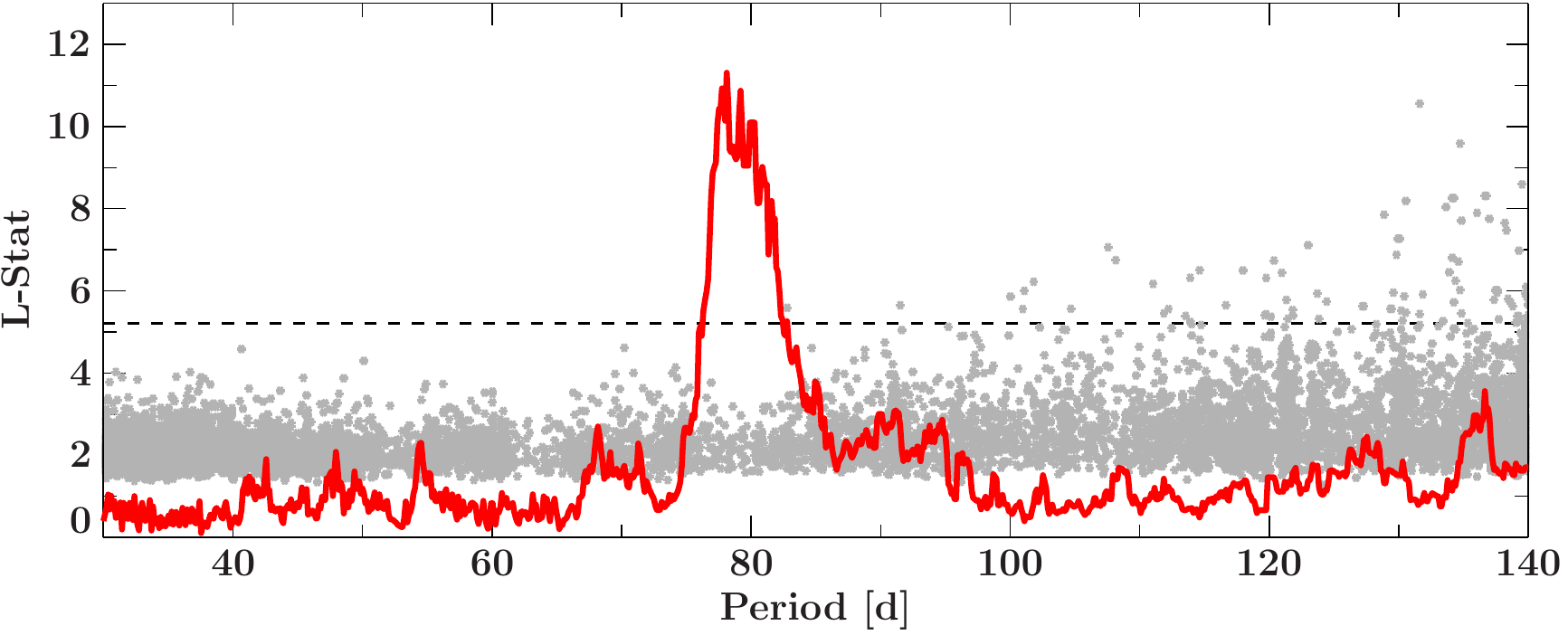}
\caption{
The results from our period search, using epoch-folding to test for periods in the
range 30--140d (see Section \ref{sec_mon}). The thick, red curve shows the
L-stat curve for the real data; a clear, strong peak is seen at $\sim$80d. The
dashed horizonal line shows the 99.9\% detection level according to the F-test
outlined in \cite{Davies90}. However, such tests often underpredict the detection
significance, so to rigorously test the statistical significance of this period we
performed a series of red-noise simulations (see Section \ref{sec_sims}). The
gray points show the peak L-stat for each of the 10,000 simulations run. None
of these reach the level of the improvement seen in the real data.
}
\vspace{0.2cm}
\label{fig_search}
\end{figure*}

We then applied the same analysis as described above to each of these simulated
datasets in order to assess the chance probability of aperiodic red-noise variability
artificially producing an apparent periodicity similar to that observed. By taking this
approach, the number of periods tested with the real data is fully accounted for in 
our estimation of the detection significance. Out of the \nsimsa\ simulations run,
\nchance\ resulted in the apparent detection of a periodicity on any of the
time-scales considered equal to or greater than that observed in the real data
(considering the full lightcurve, i.e. $L$=11.3). Figure \ref{fig_search} also shows
the maximum L-stat obtained for each of the 10,000 simulated datasets, which are
all below that seen in the real data. In order to test for any dependence on the
assumed form of the power spectrum (as discussed by \eg\ \citealt{Vaughan16} for
the case of quasar lightcurves), we also performed additional sets of simulations
varying $\alpha$ by $\pm$0.5, and obtain identical results in both cases. We
therefore conclude that the significance of the detection is comfortably in excess of
the 3$\sigma$ level.

% and likely even greater than 99.9\% significance.

\section{Period Stability}

To estimate the uncertainty of the measured period, we perform a second set of
\nsimsb\ simulations, using a similar approach to that outlined above. Here we
assume that the variability is dominated by the best-fit period following the method
outlined by \citet{Larsson96}. The same period search is again applied, and the
uncertainty determined from the distribution of the periods obtained. We find that
the variations observed have a period of $78.1\pm0.5$d (\errsig$\sigma$ error). In
addition, we compile the distribution of the widths (formally the Full-Width
Half-Maxium; FWHM) of the peak in the L-stat curves for each of these simulated
datasets in order to assess the stability of the period following the visual indication
that it may have drifted slightly during the central portion of our campaign. The
FWHM of the peak in the real data is 5.8\,d, and the expected FWHM from the
simulations assuming a stable period is $5.0\pm0.6$\,d. Despite this visual
indication, the current data are fully consistent with a stable period.

\begin{figure*}
\hspace*{-0.6cm}
\epsscale{1.13}
\plotone{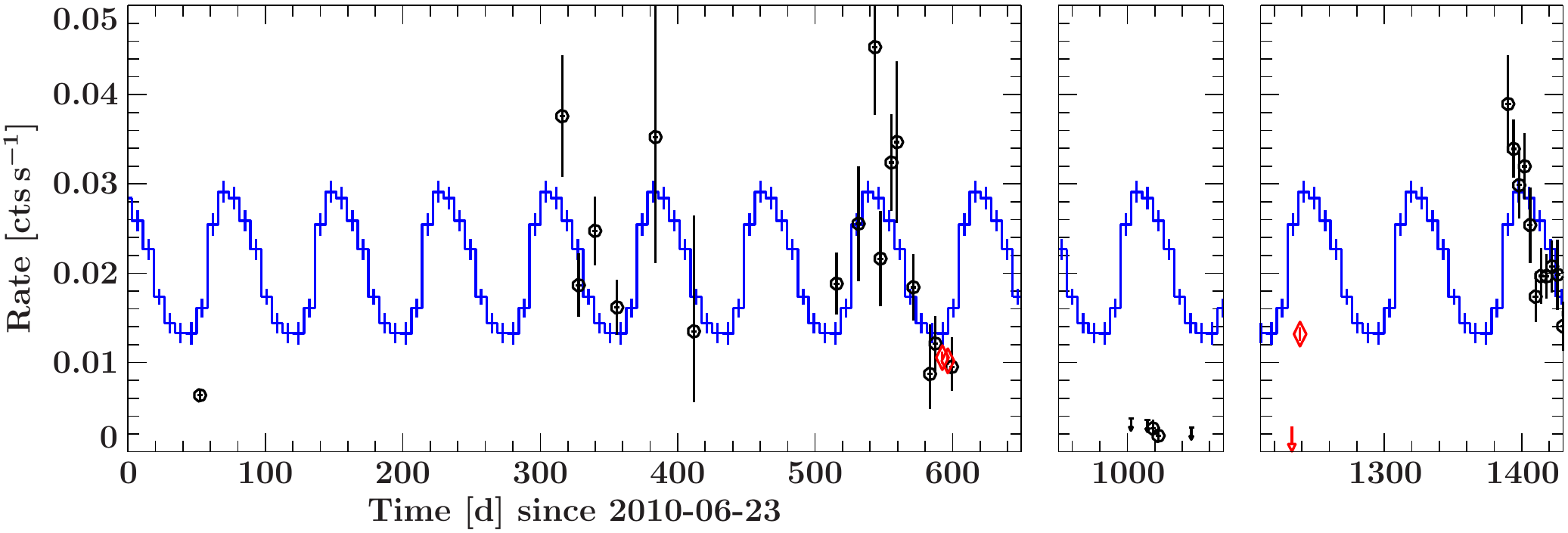}
\caption{
The best-fit cycle profile extrapolated back to the archival X-ray data obtained
in 2010--2012 (\textit{left panel}), early 2013 (centre panel) and late 2013
(\textit{right panel}). \swift\ observations are shown in black, and \xmm\
observations in red. The periodic behaviour extrapolates reasonably well to the
2010--2012 data, but can not match the 2013 \xmm\ data owing to a period of
extreme low-flux seen during this time (\citealt{Walton15}), which appears to
require a second variability mechanism.
}
\vspace{0.2cm}
\label{fig_oldobs}
\end{figure*}

As a further test of its stability we also extrapolate the detected period back to
some of the archival X-ray observations of \ngc. \xmm\ and \chandra\ both
performed observations in early 2012 (see \citealt{Sutton13}), \swift\ performed a
series of snapshots throughout 2010--2013 prior to the commencement of our
more sustained program, and, as discussed above, \xmm\ and \nustar\ also
performed two coordinated observations in late 2013 (see \citealt{Walton15}).
The extrapolation to these observations is shown in Figure \ref{fig_oldobs}.
Although the sampling is sparse, the periodic behaviour reported here extends
to the observations taken across 2010--2012, adding further support to the
stability of the period. However, the extrapolation does not match the 2013 data
well, owing to the extreme low flux states seen during this period. In particular,
the first \xmm+\nustar\ observation resulted in a non-detection of \ngc, with an
upper limit placing the flux of the source at least a factor of $\sim$50 below any
of the cycle troughs shown in Figure \ref{fig_lc} (\citealt{Walton15}). This low-flux
behaviour does not simply appear to be an extreme manifestation of the period
cycle, as the \xmm+\nustar\ observations should have occurred close to a cycle
peak. Furthermore, the series of \swift\ observations taken earlier in 2013
spanned a period of $\sim$6 weeks, and also found a systematically low flux,
$\sim$10 times lower than these cycle minima. Rather, it appears that a further
variability mechanism may have caused an extended period of low-flux during
much of 2013, and the second \xmm+\nustar\ observation in late 2013, in which
the source is well detected, caught the source on its return to its high-flux state,
in which the source is bright enough that we can detect this periodic behaviour.

For completeness, we note that in addition to these archival data, \xmm\ also
observed NGC\,5907 in 2003 (\citealt{WaltonULXcat, Sutton13}). However, while
the parameters for the period are relatively well constrained, their statistical
uncertainties are still sufficiently large to prevent a meaningful extrapolation all
the way back to these \xmm\ observations.

%\pagebreak
\section{Potential Dipping Behaviour}

There is also a hint that \ngc\ may exhibit brief dips in its observed intensity, imprinted
on top of the periodic variability. The clearest example is at day $\sim$280 in Figure
\ref{fig_lc}, where the flux is well constrained and significantly lower than both the
prediction from the average cycle profile, and also than each of the neighbouring flux
measurements. Another potential dip also appears to be seen two cycles later (at day
$\sim$440). Should these dips be real (rather than just random variability), this
behaviour could be contributing to the visual indication that the peak of the fifth cycle
arrives slightly late. There is a gap in our coverage during the rise of this cycle, and
should the coverage re-commence during one of these dips, this would give the
appearance that the peak is delayed. We note that should this be the case, the dip
would have occurred at the same phase as those seen in the cycles immediately
preceeding and following this one.

\section{Discussion and Conclusions}
\label{sec_dis}

We have reported the detection of an $78.1\pm0.5$ day X-ray periodicity in the
extreme ULX \ngc\ ($L_{\rm{X,peak}}\sim5\times10^{40}$\,\ergps) with \swift. The
variation on this timescale is very strong, with the observed XRT count rate varying
by a factor of $\sim$3--4 (peak to trough; Figure \ref{fig_lc}). Our simulations find
that this periodicity is significant, comfortably in excess of the 3$\sigma$ level.
When bright ($L_{\rm{X}}\gtrsim\times10^{40}$\,\ergps), \ngc\ dominates
the X-ray emission from NGC\,5907, so the risk of source confusion is negligible.

Long-timescale periodicities have been observed from several other well studied
ULXs. The $\sim$62 day period observed from the M82 field is well established,
generally assumed to arise from M82\,X-1 (\citealt{Kaaret07, Pasham13m82},
although this origin has recently been questioned, \citealt{Qiu15}), and a $\sim$115
day period has been claimed from NGC\,5408 X-1 (\citealt{Pasham13_5408};
although see \citealt{Grise13}). Perhaps most famously, the most luminous
ULX known to date, ESO\,243-49 HLX-1, is seen to outburst every $\sim$380
days (\citealt{Godet14}, although recently this behaviour has appeared more
erratic; \citealt{Yan15}). The timescale of the periodicity reported here is
comparable to several of these other cases.

The key question regarding the nature of the observed periodicity is whether it
could be related to the orbital period of the system, as suggested by \cite{Godet14}
for ESO\,243-49 HLX1, or perhaps some super-orbital period, as suggested by
\cite{Pasham13m82} for M82 following the likely detection of a sudden phase
shift in the cycle. Orbital periods can be imprinted on the observed lightcurves from
X-ray binaries through (at least partial) eclipses by the companion star, or through
some Be/X-ray binary-like phenomenon, in which the binary orbit is eccentric and
the accretion rate is enhanced around periastron. We note, however, that the
FRED-like cycle profile observed here does not bear much similarity to the majority
of the orbital profiles compiled by \cite{Falanga15} for eclipsing X-ray binaries. The
lightcurve observed here also does not show a series of
quiescence--outburst--quiescence cycles, as traditionally seen from Be/X ray
binaries (\citealt{Reig11}), and also from ESO\,243-49 HLX-1, so any analogy here
is limited. However, it may be possible for an elliptical orbit to result in more
moderate accretion rate variations via changes in the degree of Roche-lobe
overflow as the distance between the compact object and its stellar companion
varies (\eg\ \citealt{Church09}). Alternatively, if \ngc\ is a wind-fed X-ray binary,
the accretor could enter a higher density region of the stellar wind, resulting in
an enhanced accretion rate, similar to the case of GX\,301$-$2 (\citealt{Fuerst11}).
However, sustaining the extreme luminosities observed from \ngc\ would be a
major challenge for a wind-fed scenario.

Super-orbital X-ray periods are seen in many well monitored Galactic X-ray
binaries, \eg\ Cygnus X-1, Hercules X-1, SS433, \etc\ (\citealt{Rico08, Staubert13,
Cherepashchuk13}, and references therein), and are typically assumed to be
related to precession of the accretion flow analogous to that seen in SS433, for
which this interpretation is well established (\citealt{Fabrika04}). However,
super-orbital periods have also now been seen in wind-fed high-mass X-ray
binaries, for which such a scenario is unlikely to be viable (\citealt{Corbet13}), and
other, more exotic mechanisms such as triple systems have been proposed in
some cases (see \citealt{Kotze12} for a recent review of super-orbital variability in
X-ray binaries).

It is difficult to distinguish between these scenarios based on the observed
timescale. Several authors have suggested that even if ULXs host standard stellar
remnant black holes (\mbh\ $\sim$ 10\,\msun), some may have very long orbital
periods (up to $\sim$100 days or more) if they accrete from evolved stellar
companions via Roche-lobe overflow (\eg\ \citealt{Madhusudhan08}). Currently we
have no independent observational constraints on the nature of the stellar
companion in \ngc\ owing to both its distance and the obscuring column towards
this source ($N_{\rm{H}}\sim10^{22}$\,cm$^{-2}$; the host galaxy NGC\,5907 is
seen edge-on). However, \cite{Heida14, Heida15} have recently reported a number
of ULXs with candidate red supergiant companions, demonstrating that some of
the ULX population likely do have evolved counterparts. Indeed, if we
assume Roche-lobe overflow and that the period is orbital, we can estimate a
density for the stellar counterpart of $\rho\sim3\times10^{-5}$\,\gpcm3\
(\citealt{Faulkner72}), implying the counterpart may be either an M giant or an F
supergiant (\citealt{Drilling00}). Furthermore, the ULX \p13\ has an orbital period
of $\sim$64\,d (\citealt{Motch14nat}), so a $\sim$\period\ day orbital period may
be a plausible scenario for \ngc. Similarly, super-orbital periods have been
observed across a very wide range of timescales in Galactic systems, at least
from 3--300 days (\citealt{Kotze12}), fully consistent with the period observed here.
Should this be the correct interpretation, this would obviously imply a significantly
shorter orbital period for this system. 

In addition to the flux variations observed, we also investigated briefly whether
there is any evolution in the hardness ratio between the 0.3--2 and 2--10\,\kev\
energy bands with phase that might indicate spectral changes across the observed
cycle. We did not find any strong evidence for such variations, indicating that either
the spectrum is not systematically varying across the cycle, or that the spectral
changes are subtle enough that they are not well probed by a simple hardness ratio.
A detailed multi-epoch spectral analysis of the high S/N data available for \ngc\ will
be presented in a follow-up paper (Fuerst et al. in preparation). Ultimately, we
conclude that despite some of the orbital scenarios seeming unlikely, the question
regarding the nature of this periodicity currently remains open.

Finally, should \ngc\ be exhibiting dipping behaviour in addition to its periodic
variability, this would be of particular interest. X-ray dips have only been reported
from a handful of other ULXs to date, notably NGC\,55 ULX (\citealt{Stobbart04}),
NGC\,5408 X-1 (\citealt{Pasham13_5408, Grise13}), a source in M72 (\citealt{Lin13})
and an ultrasoft source in M51 (\citealt{Urquhart16}). Analogy with the dipping
phenomenon seen in Galactic X-ray binaries would imply we are viewing \ngc\ at a
high inclination (\eg\ \citealt{DiazTrigo06}). This would naively appear to be at odds
with the expectation from the inclination-based framework proposed to explain ULXs
with soft and hard spectra (as observed below 10\,keV) within a super-Eddington
framework, discussed in \cite{Sutton13uls} and \cite{Middleton15}. This assumes the
accretion flow has a large scale height, as expected for super-Eddington accretion
(\eg\ \citealt{Poutanen07}), resulting in an inclination dependence for the observed
X-ray spectrum. ULXs with soft spectra (as seen from NGC\,55 ULX, NGC\,5408 X-1
and the M51 source) are viewed at high inclination, such that the lower temperature
regions of the outer accretion flow dominate the observed emission and the hotter
regions of the inner flow are obscured. ULXs with hard spectra are viewed more
face on, with the hotter regions being visible. \ngc\ has a hard spectrum
(classified as a `hard ultraluminous state' by \citealt{Sutton13uls}), and so
would be expected to be viewed at a low inclination. However, it may still be
possible to reconcile dipping and a hard spectrum within this framework if our
viewing angle lies close to the opening angle of the accretion flow, such that we are
viewing the innermost regions through the uppermost atmosphere of the outer
regions, which super-Eddington similations predict to be dominated by a clumpy
outflow (\citealt{Takeuchi13}). If \ngc\ is a standard $\sim$10\,\msun\ stellar remnant,
its extreme luminosity would suggest the opening funnel for the accretion flow would
likely be quite narrow, and so the wind could well be close to our line-of-sight.

Continued monitoring of this remarkable source will test the stability of this period
over a longer baseline, helping to distinguish between orbital and super-orbital
scenarios, and may identify additional potential dips for further investigation.

\section*{ACKNOWLEDGEMENTS}

The authors would like to thank the reviewer for their timely and positive feedback,
which helped to improve the final manuscript. The work of DJW/DS was performed at
JPL/Caltech, under contract with NASA. DB and NW acknowledge financial support
from the French Space Agency (CNES), MJM acknowledges support from an Ernest
Rutherford STFC fellowship, and TPR acknowledges support from the STFC
consolidated grant ST/L00075X/1. This work made use of data supplied by the UK
Swift Science Data Centre at the University of Leicester, and also made use of the
XRT Data Analysis Software (XRTDAS) developed under the responsibility of the ASI
Science Data Center (ASDC), Italy. We acknowledge the use of public data from the
\swift\ data archive. This research has also made use of a collection of ISIS functions
(\textit{ISISscripts}) provided by ECAP/Remeis observatory and
MIT\footnote{http://www.sternwarte.uni-erlangen.de/isis/}. 

{\it Facilities:} \facility{Swift}

\bibliographystyle{/Users/dwalton/papers/mnras}

\bibliography{/Users/dwalton/papers/references}

\begin{thebibliography}{50}
\expandafter\ifx\csname natexlab\endcsname\relax\def\natexlab#1{#1}\fi

\bibitem[{Bachetti} et~al.(2013){Bachetti}, {Rana}, {Walton}
  et~al.]{Bachetti13}
{Bachetti} M., {Rana} V., {Walton} D.~J., et~al., 2013, \apj, 778, 163

\bibitem[{Burrows} et~al.(2005){Burrows}, {Hill}, {Nousek} et~al.]{SWIFT_XRT}
{Burrows} D.~N., {Hill} J.~E., {Nousek} J.~A., et~al., 2005, Space Science
  Reviews, 120, 165

\bibitem[{Cherepashchuk} et~al.(2013){Cherepashchuk}, {Sunyaev}, {Molkov},
  {Antokhina}, {Postnov} \& {Bogomazov}]{Cherepashchuk13}
{Cherepashchuk} A.~M., {Sunyaev} R.~A., {Molkov} S.~V., {Antokhina} E.~A.,
  {Postnov} K.~A., {Bogomazov} A.~I., 2013, \mnras, 436, 2004

\bibitem[{Church} et~al.(2009){Church}, {Dischler}, {Davies}, {Tout}, {Adams}
  \& {Beer}]{Church09}
{Church} R.~P., {Dischler} J., {Davies} M.~B., {Tout} C.~A., {Adams} T., {Beer}
  M.~E., 2009, \mnras, 395, 1127

\bibitem[{Corbet} \& {Krimm}(2013)]{Corbet13}
{Corbet} R.~H.~D., {Krimm} H.~A., 2013, \apj, 778, 45

\bibitem[{Davies}(1990)]{Davies90}
{Davies} S.~R., 1990, \mnras, 244, 93

\bibitem[{D{\'{\i}}az Trigo} et~al.(2006){D{\'{\i}}az Trigo}, {Parmar},
  {Boirin}, {M{\'e}ndez} \& {Kaastra}]{DiazTrigo06}
{D{\'{\i}}az Trigo} M., {Parmar} A.~N., {Boirin} L., {M{\'e}ndez} M., {Kaastra}
  J.~S., 2006, \aap, 445, 179

\bibitem[{Drilling} \& {Landolt}(2000)]{Drilling00}
{Drilling} J.~S., {Landolt} A.~U., 2000, {Normal Stars},  381

\bibitem[{Evans} et~al.(2009){Evans}, {Beardmore}, {Page} et~al.]{Evans09}
{Evans} P.~A., {Beardmore} A.~P., {Page} K.~L., et~al., 2009, \mnras, 397, 1177

\bibitem[{Fabrika}(2004)]{Fabrika04}
{Fabrika} S., 2004, Astrophysics and Space Physics Reviews, 12, 1

\bibitem[{Falanga} et~al.(2015){Falanga}, {Bozzo}, {Lutovinov},
  {Bonnet-Bidaud}, {Fetisova} \& {Puls}]{Falanga15}
{Falanga} M., {Bozzo} E., {Lutovinov} A., {Bonnet-Bidaud} J.~M., {Fetisova} Y.,
  {Puls} J., 2015, \aap, 577, A130

\bibitem[{Faulkner} et~al.(1972){Faulkner}, {Flannery} \& {Warner}]{Faulkner72}
{Faulkner} J., {Flannery} B.~P., {Warner} B., 1972, \apjl, 175, L79

\bibitem[{F{\"u}rst} et~al.(2011){F{\"u}rst}, {Suchy}, {Kreykenbohm}
  et~al.]{Fuerst11}
{F{\"u}rst} F., {Suchy} S., {Kreykenbohm} I., et~al., 2011, \aap, 535, A9

\bibitem[{Gehrels} et~al.(2004){Gehrels}, {Chincarini}, {Giommi} et~al.]{SWIFT}
{Gehrels} N., {Chincarini} G., {Giommi} P., et~al., 2004, \apj, 611, 1005

\bibitem[{Godet} et~al.(2014){Godet}, {Lombardi}, {Antonini} et~al.]{Godet14}
{Godet} O., {Lombardi} J.~C., {Antonini} F., et~al., 2014, \apj, 793, 105

\bibitem[{Gris{\'e}} et~al.(2013){Gris{\'e}}, {Kaaret}, {Corbel}, {Cseh} \&
  {Feng}]{Grise13}
{Gris{\'e}} F., {Kaaret} P., {Corbel} S., {Cseh} D., {Feng} H., 2013, \mnras,
  433, 1023

\bibitem[{Heida} et~al.(2014){Heida}, {Jonker}, {Torres} et~al.]{Heida14}
{Heida} M., {Jonker} P.~G., {Torres} M.~A.~P., et~al., 2014, \mnras, 442, 1054

\bibitem[{Heida} et~al.(2015){Heida}, {Torres}, {Jonker} et~al.]{Heida15}
{Heida} M., {Torres} M.~A.~P., {Jonker} P.~G., et~al., 2015, \mnras, 453, 3511

\bibitem[{Kaaret} \& {Feng}(2007)]{Kaaret07}
{Kaaret} P., {Feng} H., 2007, \apj, 669, 106

\bibitem[{Kotze} \& {Charles}(2012)]{Kotze12}
{Kotze} M.~M., {Charles} P.~A., 2012, \mnras, 420, 1575

\bibitem[{Larsson}(1996)]{Larsson96}
{Larsson} S., 1996, \aaps, 117, 197

\bibitem[{Leahy}(1987)]{Leahy87}
{Leahy} D.~A., 1987, \aap, 180, 275

\bibitem[{Lin} et~al.(2013){Lin}, {Irwin}, {Webb}, {Barret} \&
  {Remillard}]{Lin13}
{Lin} D., {Irwin} J.~A., {Webb} N.~A., {Barret} D., {Remillard} R.~A., 2013,
  \apj, 779, 149

\bibitem[{Madhusudhan} et~al.(2008){Madhusudhan}, {Rappaport}, {Podsiadlowski}
  \& {Nelson}]{Madhusudhan08}
{Madhusudhan} N., {Rappaport} S., {Podsiadlowski} P., {Nelson} L., 2008, \apj,
  688, 1235

\bibitem[{Middleton} et~al.(2015){Middleton}, {Heil}, {Pintore}, {Walton} \&
  {Roberts}]{Middleton15}
{Middleton} M.~J., {Heil} L., {Pintore} F., {Walton} D.~J., {Roberts} T.~P.,
  2015, \mnras, 447, 3243

\bibitem[{Motch} et~al.(2014){Motch}, {Pakull}, {Soria}, {Gris{\'e}} \&
  {Pietrzy{\'n}ski}]{Motch14nat}
{Motch} C., {Pakull} M.~W., {Soria} R., {Gris{\'e}} F., {Pietrzy{\'n}ski} G.,
  2014, \nat, 514, 198

\bibitem[{Mukherjee} et~al.(2015){Mukherjee}, {Walton}, {Bachetti}
  et~al.]{Mukherjee15}
{Mukherjee} E.~S., {Walton} D.~J., {Bachetti} M., et~al., 2015, \apj, 808, 64

\bibitem[{Pasham} \& {Strohmayer}(2013{\natexlab{a}})]{Pasham13m82}
{Pasham} D.~R., {Strohmayer} T.~E., 2013{\natexlab{a}}, \apjl, 774, L16

\bibitem[{Pasham} \& {Strohmayer}(2013{\natexlab{b}})]{Pasham13_5408}
{Pasham} D.~R., {Strohmayer} T.~E., 2013{\natexlab{b}}, \apj, 764, 93

\bibitem[{Poutanen} et~al.(2007){Poutanen}, {Lipunova}, {Fabrika}, {Butkevich}
  \& {Abolmasov}]{Poutanen07}
{Poutanen} J., {Lipunova} G., {Fabrika} S., {Butkevich} A.~G., {Abolmasov} P.,
  2007, \mnras, 377, 1187

\bibitem[{Qiu} et~al.(2015){Qiu}, {Liu}, {Guo} \& {Wang}]{Qiu15}
{Qiu} Y., {Liu} J., {Guo} J., {Wang} J., 2015, \apjl, 809, L28

\bibitem[{Rana} et~al.(2015){Rana}, {Harrison}, {Bachetti} et~al.]{Rana15}
{Rana} V., {Harrison} F.~A., {Bachetti} M., et~al., 2015, \apj, 799, 121

\bibitem[{Reig}(2011)]{Reig11}
{Reig} P., 2011, \apss, 332, 1

\bibitem[{Remillard} \& {McClintock}(2006)]{Remillard06rev}
{Remillard} R.~A., {McClintock} J.~E., 2006, \araa, 44, 49

\bibitem[{Rico}(2008)]{Rico08}
{Rico} J., 2008, \apjl, 683, L55

\bibitem[{Staubert} et~al.(2013){Staubert}, {Klochkov}, {Vasco}
  et~al.]{Staubert13}
{Staubert} R., {Klochkov} D., {Vasco} D., et~al., 2013, \aap, 550, A110

\bibitem[{Stobbart} et~al.(2004){Stobbart}, {Roberts} \& {Warwick}]{Stobbart04}
{Stobbart} A.-M., {Roberts} T.~P., {Warwick} R.~S., 2004, \mnras, 351, 1063

\bibitem[{Sutton} et~al.(2013{\natexlab{a}}){Sutton}, {Roberts}, {Gladstone}
  et~al.]{Sutton13}
{Sutton} A.~D., {Roberts} T.~P., {Gladstone} J.~C., et~al., 2013{\natexlab{a}},
  \mnras, 434, 1702

\bibitem[{Sutton} et~al.(2013{\natexlab{b}}){Sutton}, {Roberts} \&
  {Middleton}]{Sutton13uls}
{Sutton} A.~D., {Roberts} T.~P., {Middleton} M.~J., 2013{\natexlab{b}}, \mnras,
  435, 1758

\bibitem[{Sutton} et~al.(2012){Sutton}, {Roberts}, {Walton}, {Gladstone} \&
  {Scott}]{Sutton12}
{Sutton} A.~D., {Roberts} T.~P., {Walton} D.~J., {Gladstone} J.~C., {Scott}
  A.~E., 2012, \mnras, 423, 1154

\bibitem[{Takeuchi} et~al.(2013){Takeuchi}, {Ohsuga} \&
  {Mineshige}]{Takeuchi13}
{Takeuchi} S., {Ohsuga} K., {Mineshige} S., 2013, \pasj, 65

\bibitem[{Timmer} \& {Koenig}(1995)]{Timmer95}
{Timmer} J., {Koenig} M., 1995, \aap, 300, 707

\bibitem[{Urquhart} \& {Soria}(2016)]{Urquhart16}
{Urquhart} R., {Soria} R., 2016, \mnras, 456, 1859

\bibitem[{Vaughan} et~al.(2003){Vaughan}, {Fabian} \& {Nandra}]{Vaughan03}
{Vaughan} S., {Fabian} A.~C., {Nandra} K., 2003, \mnras, 339, 1237

\bibitem[{Vaughan} et~al.(2016){Vaughan}, {Uttley}, {Markowitz}
  et~al.]{Vaughan16}
{Vaughan} S., {Uttley} P., {Markowitz} A.~G., et~al., 2016, ArXiv e-prints

\bibitem[{Walton} et~al.(2015{\natexlab{a}}){Walton}, {Harrison}, {Bachetti}
  et~al.]{Walton15}
{Walton} D.~J., {Harrison} F.~A., {Bachetti} M., et~al., 2015{\natexlab{a}},
  \apj, 799, 122

\bibitem[{Walton} et~al.(2014){Walton}, {Harrison}, {Grefenstette}
  et~al.]{Walton14hoIX}
{Walton} D.~J., {Harrison} F.~A., {Grefenstette} B.~W., et~al., 2014, \apj,
  793, 21

\bibitem[{Walton} et~al.(2015{\natexlab{b}}){Walton}, {Middleton}, {Rana}
  et~al.]{Walton15hoII}
{Walton} D.~J., {Middleton} M.~J., {Rana} V., et~al., 2015{\natexlab{b}}, \apj,
  806, 65

\bibitem[{Walton} et~al.(2011){Walton}, {Roberts}, {Mateos} \&
  {Heard}]{WaltonULXcat}
{Walton} D.~J., {Roberts} T.~P., {Mateos} S., {Heard} V., 2011, \mnras, 416,
  1844

\bibitem[{Yan} et~al.(2015){Yan}, {Zhang}, {Soria}, {Altamirano} \&
  {Yu}]{Yan15}
{Yan} Z., {Zhang} W., {Soria} R., {Altamirano} D., {Yu} W., 2015, \apj, 811, 23

\end{thebibliography}

\label{lastpage}

\end{document}